\documentclass[12pt]{iopart}
\usepackage{graphicx}
\usepackage{subcaption}
\usepackage{float}

\begin{document}

\title[]{Single photoelectron identification with Incom LAPPD 38}

\author{S. P. Malace}
\address{Thomas Jefferson National Accelerator Facility, 12000 Jefferson Avenue, Newport News, VA 23606, USA}
\ead{simona@jlab.org}
\author{S. Wood}
\address{Thomas Jefferson National Accelerator Facility, 12000 Jefferson Avenue, Newport News, VA 23606, USA}

\vspace{10pt}
\begin{indented}
\item[]April 2021
\end{indented}

\begin{abstract}
Incom Inc. Large Area Picosecond Photodetector (LAPPD) 38 has been tested at Jefferson Lab to identify single-photoelectron signals to assess the potential of this type of device for future applications in Cherenkov light detection. Single-photoelectron signals were clearly detected if a tight masking of photons impinging on the photocathode was used compared to the pixelation of the charge collection signal board. 

\end{abstract}

%
%
%
%
%

\section{Introduction}

Large Area Picosecond Photodetectors (LAPPDs) are now available for use in high energy and nuclear physics experiments where picosecond-level timing, high spatial resolution, resistance to magnetic fields and large area, cost effective photosensor coverage is required. A recent review of the status and performance of routinely produced LAPPDs is presented in refs.~\cite{minot,lyashenko,ertley,minot2,popecki,frisch}. In this paper we explore the capability of one of the Incom Inc. produced LAPPDs to identify single photoelectron signals for possible applications in Cherenkov detectors.  

A photosensor with a pixelated response is desirable in Cherenkov detector applications where a high event rate is expected as it can allow for a significant reduction of background and, therefore, for a much needed improvement of signal to background ratio. Typically, Cherenkov cones spread over areas of the order of few cm$^{2}$ or larger so one can choose a pixelation of the photosensor that allows for significant background rejection at the trigger level or offline by requiring coincidence between several adjacent photosensor pixels. The Incom Inc. LAPPD has 200 by 200 mm Multichannel Plates (MCPs), and a position resolution on the order of a millimeter. The default pixel size on the signal board is 25 mm square. This may be changed to any pattern simply by producing a new signal board. The footprint of the MCP pulse at the anode is approximately 7 mm in diameter, which includes about 70$\%$ of the charge. Consequently, a MCP pulse may be simultaneously observed by capacitive coupling on two adjacent 25 mm pixels. Centroiding the charge leads to a position resolution of a millimeter, which is much better than the pixel size. One could also sum the charge over adjacent pixels, if the high position resolution is not needed but instead just the total charge of the MCP pulse will suffice. Lastly, if a more discrete pixelated response is needed, then a design change is required in which the anode is placed ~1 mm away from the MCP, rather than the 6.5 mm standard spacing of the LAPPD. In this initial study, a standard, capacitively-coupled LAPPD is evaluated for Cherenkov light detection, using masking on the entry window to create the desired spatially discrete response.

\section{Test Setup}

\begin{figure}[H]
  \centering
  \includegraphics[width=0.8\textwidth]{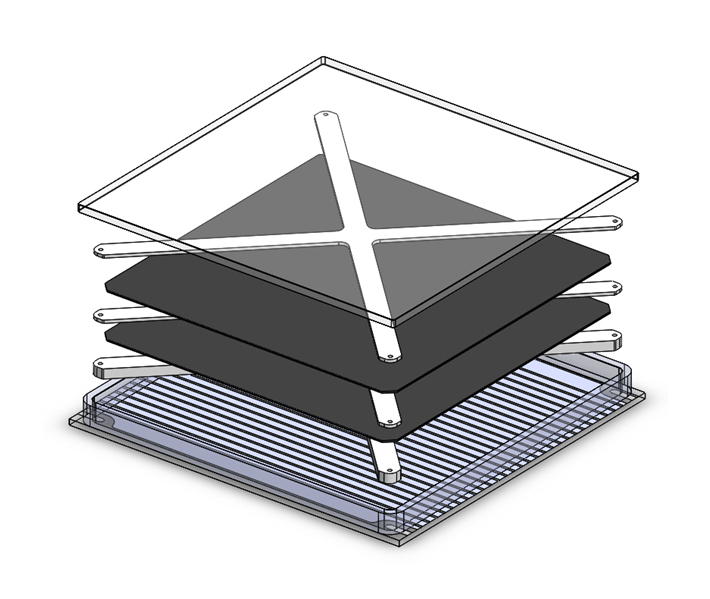}
  \caption{A schematic of the LAPPD layers described in Table~\ref{tab:table_lappd}.}
  \label{fig:lappd_disection}
\end{figure}

\begin{figure}[H]
  \centering
  \includegraphics[width=0.99\textwidth]{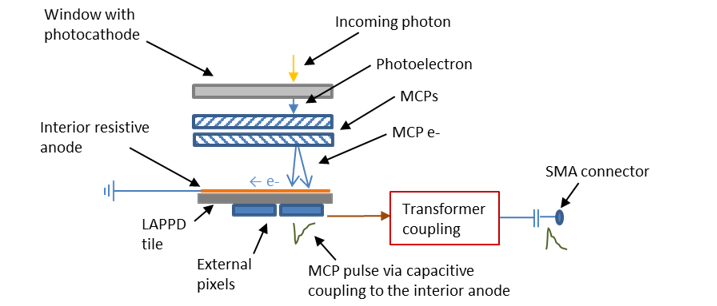}
  \caption{A cross-section of the capacitively-coupled LAPPD is shown. A photon initiates a MCP pulse. The charge in the pulse arrives at the resistive anode and propagates to ground at the edge of the anode. The external pixels capacitively detect the pulse and the pulse is brought to a connector at the edge of the signal board.}
  \label{fig:capacitive_readout}
\end{figure}

\begin{figure}[H]
  \centering
  \includegraphics[width=0.6\textwidth]{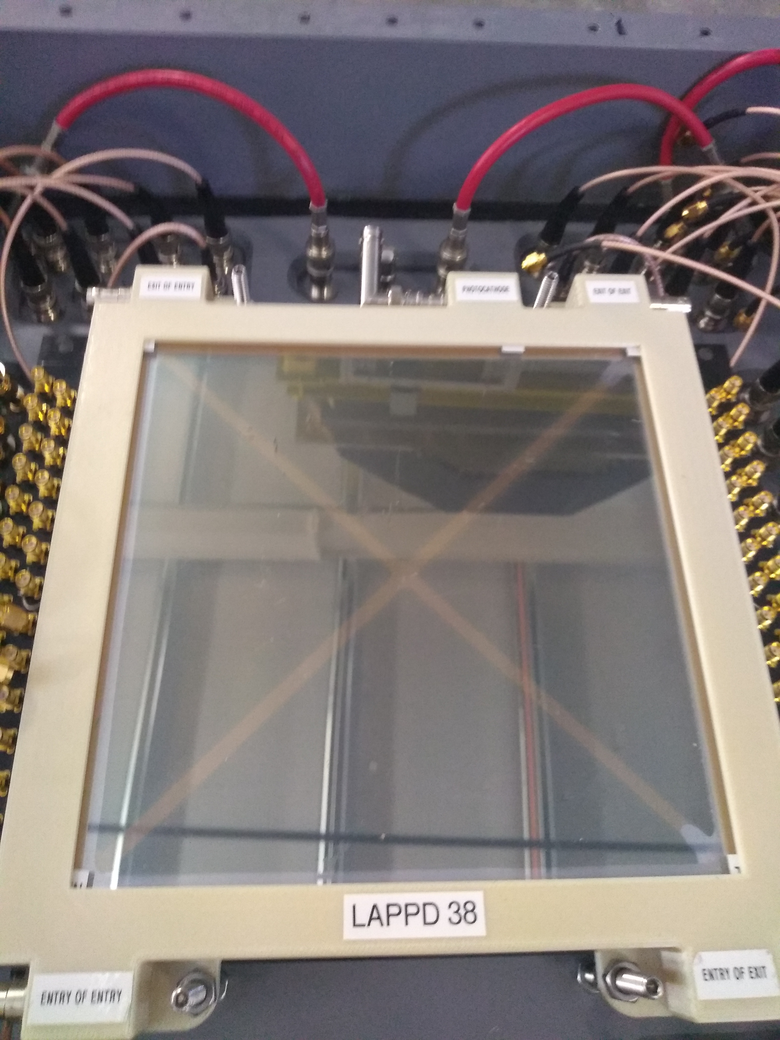}
  \caption{A picture of Incom Inc. LAPPD 38. The cross structure is an integral part of the detector and it is used to support the LAPPD planes. The face of the LAPPD as shown was exposed to photons generated by a blue LED using various masking configurations (see text for details).}
  \label{fig:lappdpic}
\end{figure}

LAPPD 38 was manufactured by Incom Inc. (12/21/2018) and was made available for bench tests at Jefferson Lab. LAPPD 38 was one of the earliest GEN II LAPPDs fabricated by Incom. The GEN II design offers capacitively coupled signal read out to pixelated anodes located on a printed circuit board that is deployed beneath and external to the evacuated photodetector tile. This design concept allows end users to customize the readout according to their needs by using a signal board with pixels of the desired size. In preparation for testing at Jefferson Lab, the LAPPD tile window (made of Borofloat glass) had been coated with a p-Terphenyl wave shifting coating, which absorbs UV photons in a 240 to 300 nm range and re-emits them in the 320 to 400 nm range.  LAPPD 38 has a quantum efficiency of 16.7$\%$ at 365 nm (meausrement taken before the wavelength shifter application) and a chevron pair of ALD-GCA-MCPs with 20-micron pore size that provide signal amplification.  The printed circuit signal board used had 64 2.5 by 2.5 cm charge collection pixels ref.~\cite{minot}. More details on the LAPPD 38 design characteristics are given in Table~\ref{tab:table_lappd} and Figures~\ref{fig:lappd_disection},~\ref{fig:capacitive_readout},~\ref{fig:lappdpic}.

\begin{table}[h!]
\begin{center}
\caption{LAPPD design characteristics}
\label{tab:table_lappd}
\begin{tabular}{l|c}
\hline
window & 5 mm thick borosilicate \\
\hline
photocathode & potassium, sodium, antimony \\
& 0.345 $\mu$m thick \\
\hline
photocathode - first MCP gap & 2.8 mm created via X-spacers \\
\hline
first MCP & borosilicate, 65$\%$ open area ratio \\
& 1.2 mm thick \\  
\hline
gap between first and second MCP & 1.1 mm created via X-spacers \\
\hline 
second MCP & borosilicate, 65$\%$ open area ratio \\
& 1.2 mm thick \\  
\hline
second MCP - anode gap & 6.6 mm created via X-spacers \\
\hline 
anode & 3.8 mm borosilicate with 12 $\mu$m thick silver strips \\
\hline
\end{tabular}
\end{center}
\end{table}

The capacitive coupling between pixels and anode may be described as follows. The incident photoelectron creates a MCP pulse. The pulse strikes the resistive anode and propagates to ground at the edge of the anode. The propagation time is longer than the typical width of an MCP pulse. The arrival of the pulses is promptly detected on a pixel located beneath the LAPPD baseplate on a signal board. This pulse is then routed to the edge of the signal board to a connector. The routing makes use of a transformer coupling to preserve a 50 ohm impedance and avoid pulse reflections. The capacitive coupling and the waveform of the pulse will be affected by the dielectric constant of the LAPPD base material, the thickness of the base, and the area of the pixel beneath the base. The capacitive coupling concept is also illustrated in Fig.~\ref{fig:capacitive_readout}.

\begin{figure}[H]
  \centering
  \includegraphics[width=0.7\textwidth]{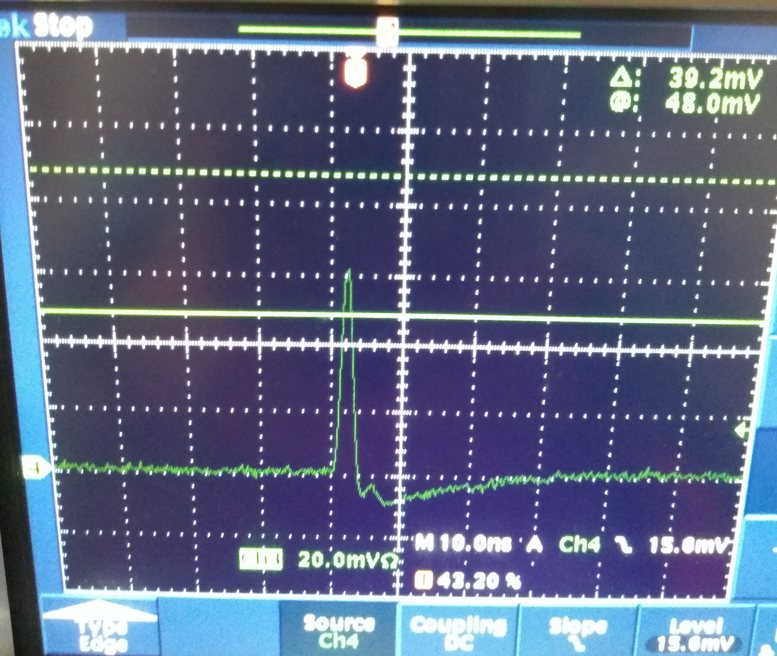}
  \caption{A typical LAPPD pulse displayed on an oscilloscope.}
  \label{fig:lappdpulseyay}
\end{figure}

\begin{figure}[H]
  \centering
    \includegraphics[width=0.7\linewidth]{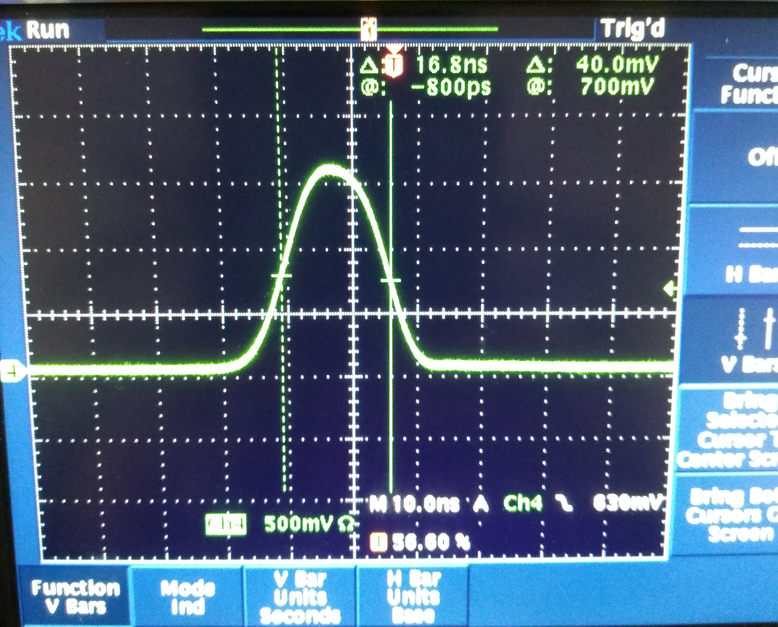}
    \caption{A typical pulse produced by the Agilent waveform generator used to drive the LED and trigger the data acquisition system. The width and the frequency of the pulse were kept fixed for the entire duration of the test. The voltage was varied to control the yield of photons emitted by the LED.}
    \label{fig:trigger}
  \end{figure}

\begin{figure}[H]
    \centering
    \includegraphics[width=0.65\linewidth]{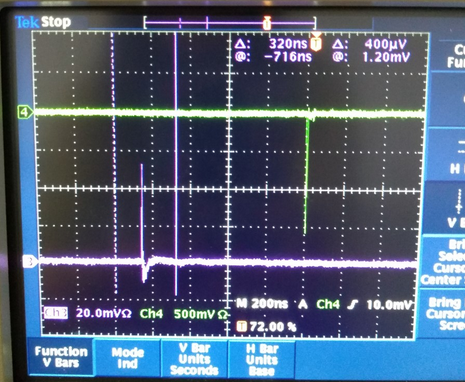}
    \caption{The LAPPD pulse is displayed in pink while the signal that triggers the data acquisition system is shown in green. The vertical lines represent the sampling window of the FADC250 module.}
    \label{fig:trigger_2}
  \end{figure}

At Jefferson Lab the LAPPD was tested in a dark box using light from a blue LED. We chose to use LED produced photons as opposed to laser light to better simulate the typical light distribution on a photosensor in a Cherenkov detector. A typical LAPPD pulse as viewed on an oscilloscope is shown in Fig.~\ref{fig:lappdpulseyay}. A Hamamatsu multi-anode PMT (maPMT) H12700 was used as witness to the LAPPD test. Both devices were placed in a dark box well insulated from ambient light and were tested at the same time. The LED was fixed to the top cover of the dark box and kept at the same location for the duration of the test. The LED was pulsed by 16 ns wide pulses of variable voltage at a rate of 500 Hz as produced by an Agilent 33522A waveform generator. The charge from the LAPPD and the multi-anode PMT was digitized by a 16-channel FADC250 module which samples every 4 ns. The data acquisition was triggered by the pulse that drove the LED. Fig.~\ref{fig:trigger} shows a typical pulse used to drive the LED while Fig.~\ref{fig:trigger_2} shows the trigger signal and LAPPD signals on an oscilloscope. 

Several test configurations were used to identify single-photoelectron signals. First one quarter of the LAPPD photocathode was exposed to LED light (the quarter was selected by masking the face of the LAPPD from light except for one quadrant), then individual pixels (2.5 by 2.5 cm areas) one at a time and finally two adjacent pixels together. When individual pixels were exposed to light a masking of $\approx$75$\%$ of each pixel's area was used (only 25$\%$ of the area corresponding to a readout pixel was exposed to light) as this yielded better resolution for single photoelectron identification. For each test configuration the LED voltage was varied in incremental steps as means of varying the probability for single-photoelectron production for both the LAPPD and the maPMT. For each LED voltage setting a run of 200,000 triggers was taken. The maPMT test results were monitored throughout the test to check the functionality of the setup.

\section{Results}

In what follows the results of the test are presented. First one example of the maPMT measurements is shown. Then the LAPPD results are presented and discussed. 

\subsection{maPMT test results}

\begin{figure}[H]
  \centering
        \vspace{-1.5\baselineskip}
  \includegraphics[width=0.85\textwidth]{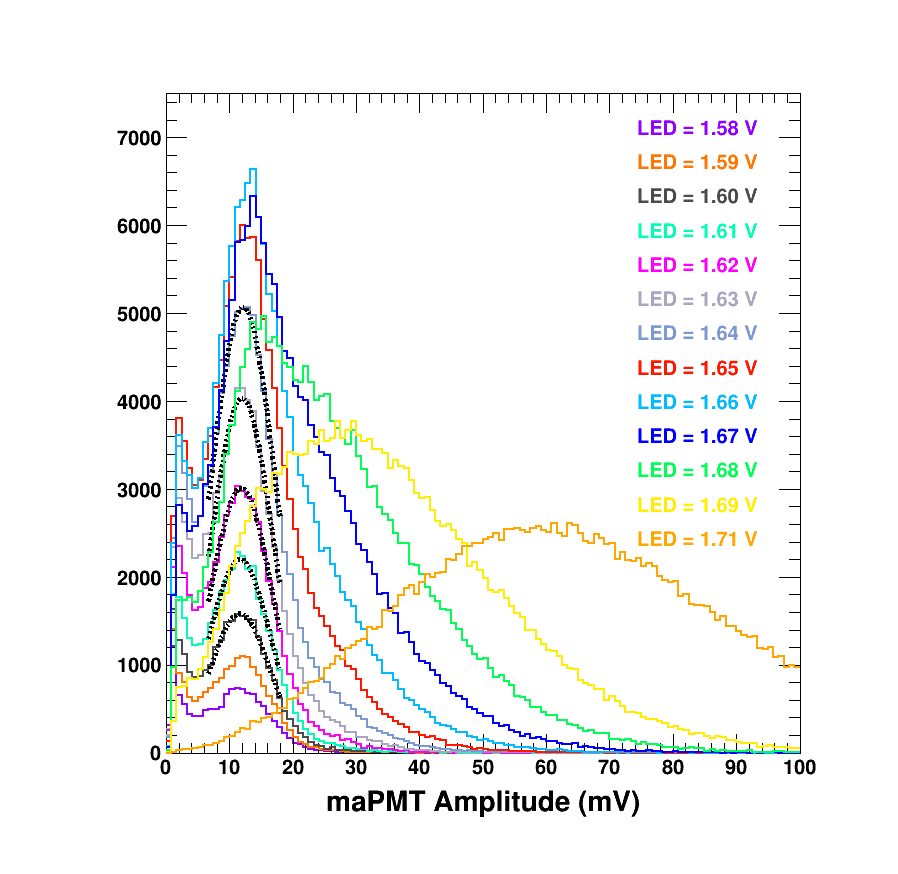}
  \caption{Amplitude distributions of single photoelectron signals from a Hamamatsu multi-anode PMT H12700 acquired with different LED voltage settings which ensures different intensities of photons impinging on the PMT's photocathode. The same number of triggers per LED settings were taken to highlight the change in probability of single photoelectron production. The dashed lines represent simple Gaussian fits to determine the single photoelectron amplitude (see text for more details).}
  \label{fig:mapmt_1}
\end{figure}

\begin{figure}[H]
  \centering
  \includegraphics[width=0.65\textwidth]{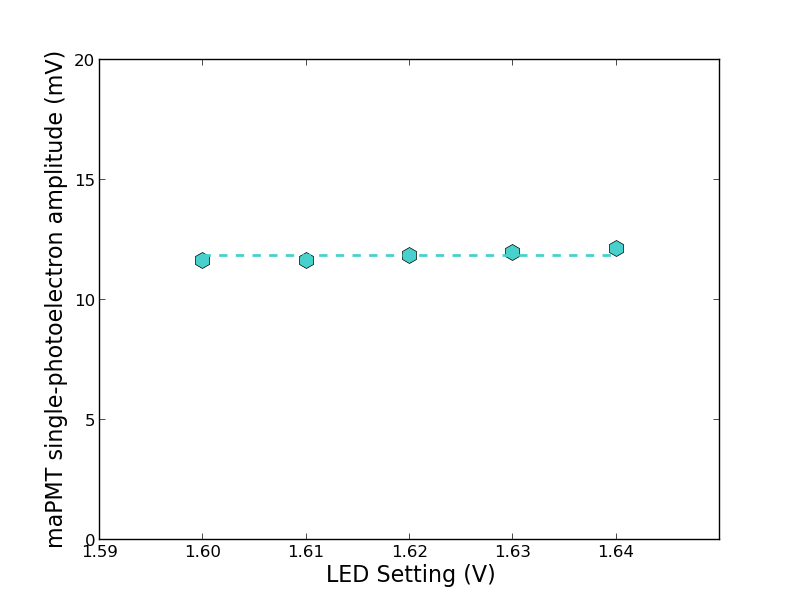}
  \caption{Summary of maPMT single photoelectron amplitudes from the Gaussian fits. The single photoelectron amplitude thus obtained is consistent with the amplitude observed on an oscilloscope.}
  \label{fig:mapmt_2}
\end{figure}

A representative plot showing amplitude distributions of signals from maPMT H12700B-03 is shown in Fig.~\ref{fig:mapmt_1}. The different amplitude distributions shown correspond to different LED voltage settings that result in a change in the number of photons impinging on the maPMT photocathode. The same number of triggers were taken to produce each one of the distributions shown in order to highlight the change in the probability of producing single or multi photoelectrons as the LED photon yield changes with the change in the voltage applied to the LED. In Fig.~\ref{fig:mapmt_1} the pulse amplitude distributions obtained when the LED was pulsed at low voltages between 1.58 and 1.64 V represent predominantly events where one single photoelectron was extracted from the photocathode: the average maPMT pulse amplitude stays the same as single photoelectrons are produced predominantly (the pulse amplitude is shown on the x axis) the only change being the rate of single photoelectron production for the same number of triggers (pulses that drive the LED). As the LED voltage was increased beyond 1.64 V and up to 1.71 V an admixture of events was produced as the probability of extracting more than one photoelectron from the photocathode became non-negligible. A summary plot displaying the maPMT single photoelectron amplitude extracted from a simple Gaussian fit to those amplitude distributions dominated by single-photoelectron events is shown in Fig.~\ref{fig:mapmt_2}. The single photoelectron amplitude thus obtained is consistent with the amplitude observed on an oscilloscope. The maPMT test results show that with the test setup described above single photoelectron events can be identified. 

\subsection{LAPPD test results}

\begin{figure}
  \centering
        \vspace{-1.5\baselineskip}
  \includegraphics[width=0.7\textwidth]{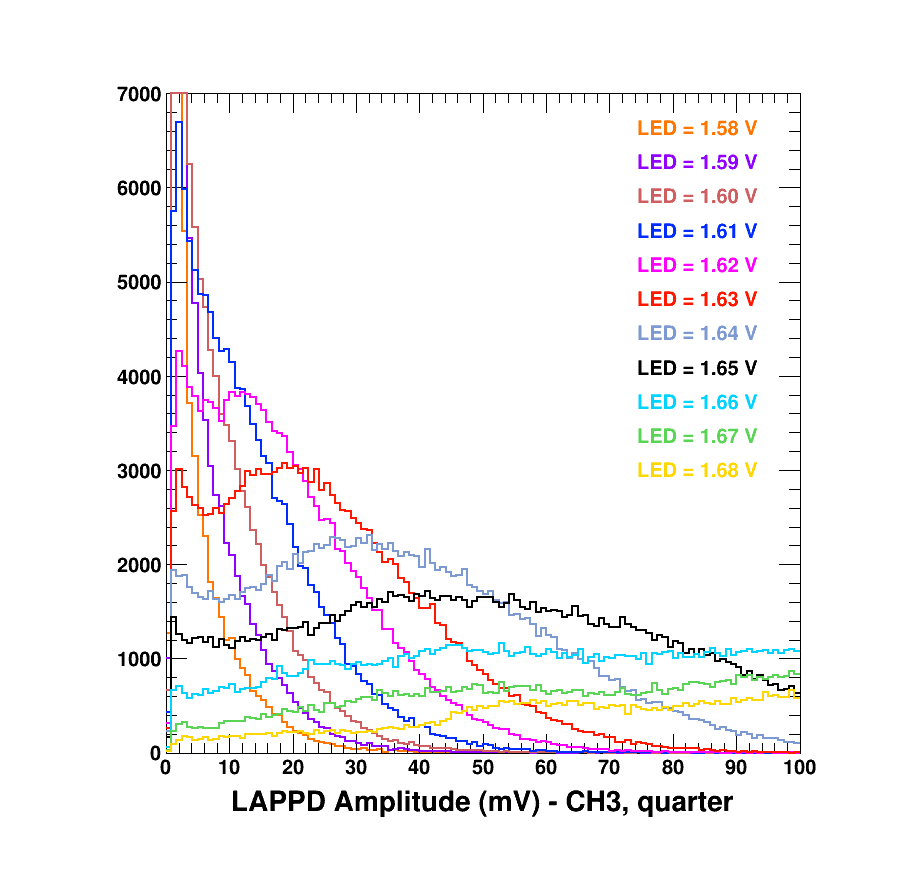}
  \caption{LAPPD CH3 (CH3 is a label for one uniquely identifiable readout pixel) signal amplitudes when a quarter of the LAPPD is illuminated. The signals were acquired from a single pixel in that quadrant. Other readout pixels yielded very similar results.}
  \label{fig:quarter}
\end{figure}

A similar test was performed with the LAPPD. First a quadrant of the LAPPD was illuminated by the LED, in steps of increasing LED voltage. Several pixels from the quadrant were monitored. In this arrangement, a given pixel can detect MCP pulses immediately above, or in adjacent locations, with correspondingly varying measured amplitudes. The pulse amplitude distributions were broader than for the maPMT as shown in Fig.~\ref{fig:quarter} and unlike for the maPMT, the signature of single-photoelectron production at the lowest LED voltages was not readily apparent. It should be noted that the maPMT has a set of discrete anodes, and photoelectrons are electrostatically steered from the photocathode to each anode to enhance the spatially localized response of the device. In the case of the LAPPD when a larger area of the photocathode is illuminated the pixel of interest will collect charge of avalanches from single photoelectrons right above it but also partially charge from single photoelectrons that hit adjacent areas to the one above it. 

\begin{figure}[H]
  \captionsetup[subfigure]{aboveskip=5pt,belowskip=5pt}
	\centering
        \vspace{-1.5\baselineskip}
	\begin{subfigure}{0.78\textwidth} 
		\includegraphics[width=\textwidth]{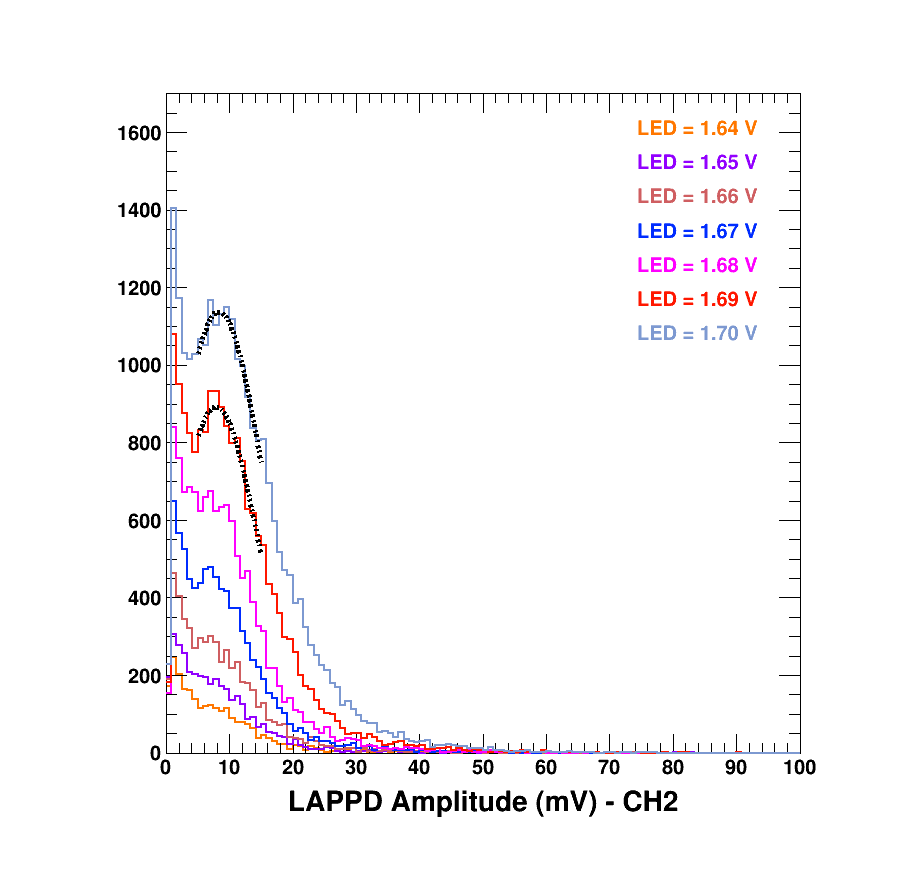}
                \vspace{-3.45\baselineskip}
	\end{subfigure}
	\begin{subfigure}{0.78\textwidth} 
		\includegraphics[width=\textwidth]{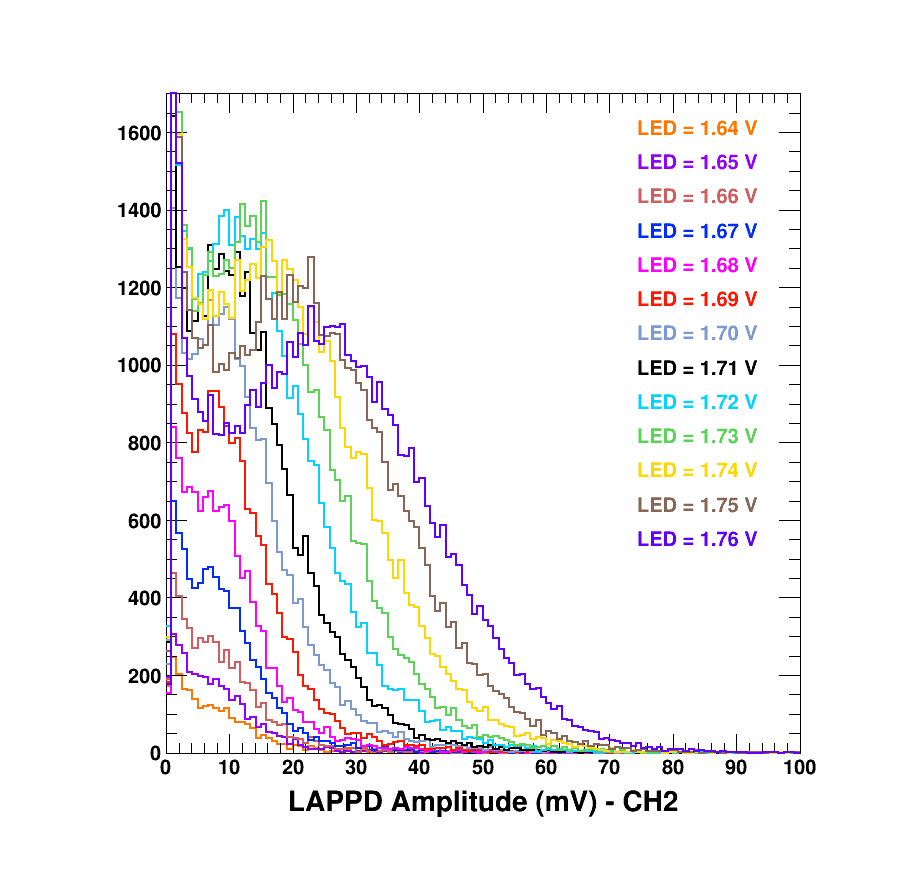}
	\end{subfigure}
        \vspace{-1.5\baselineskip}
	\caption{LAPPD CH2 (CH2 is a label for one uniquely identifiable pixel) signal amplitudes when about 25$\%$ of the LAPPD region corresponding to readout pixel CH2 is illuminated. Top: for clarity only a subset of the data is shown. Bottom: the entire data set is shown.} 
\label{fig:CH2fig}
\end{figure}

\begin{figure}[H]
  \captionsetup[subfigure]{aboveskip=5pt,belowskip=5pt}
	\centering
        \vspace{-1.5\baselineskip}
	\begin{subfigure}{0.78\textwidth} 
		\includegraphics[width=\textwidth]{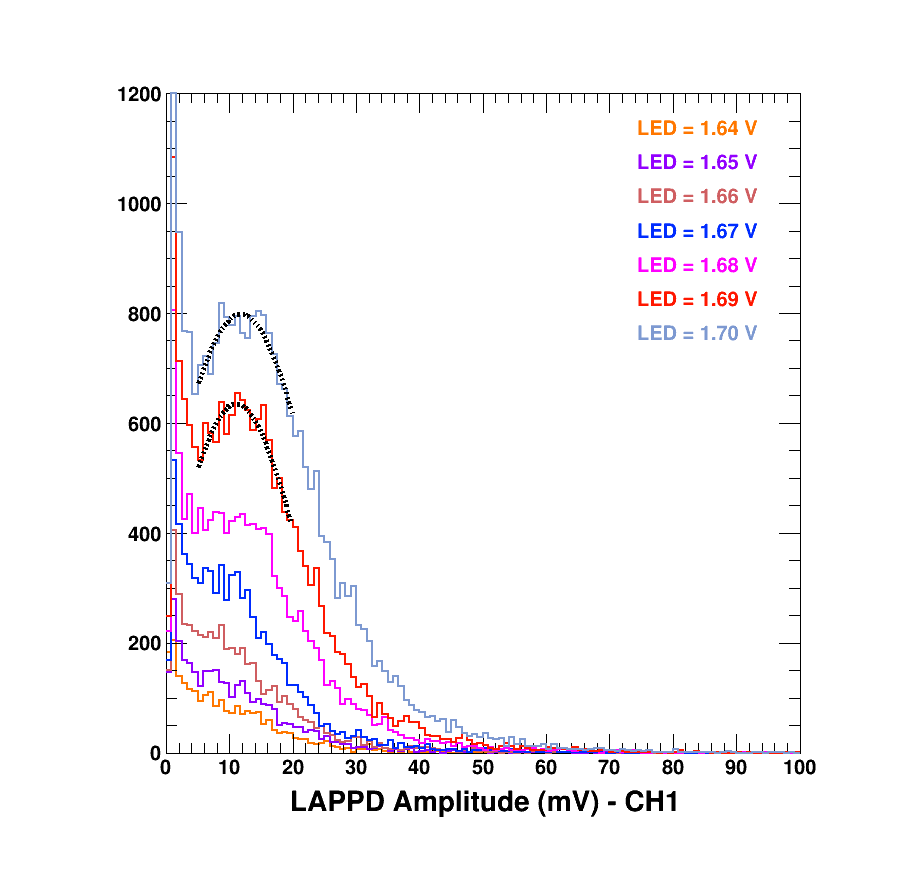}
                \vspace{-3.45\baselineskip}
	\end{subfigure}
	\begin{subfigure}{0.78\textwidth} 
		\includegraphics[width=\textwidth]{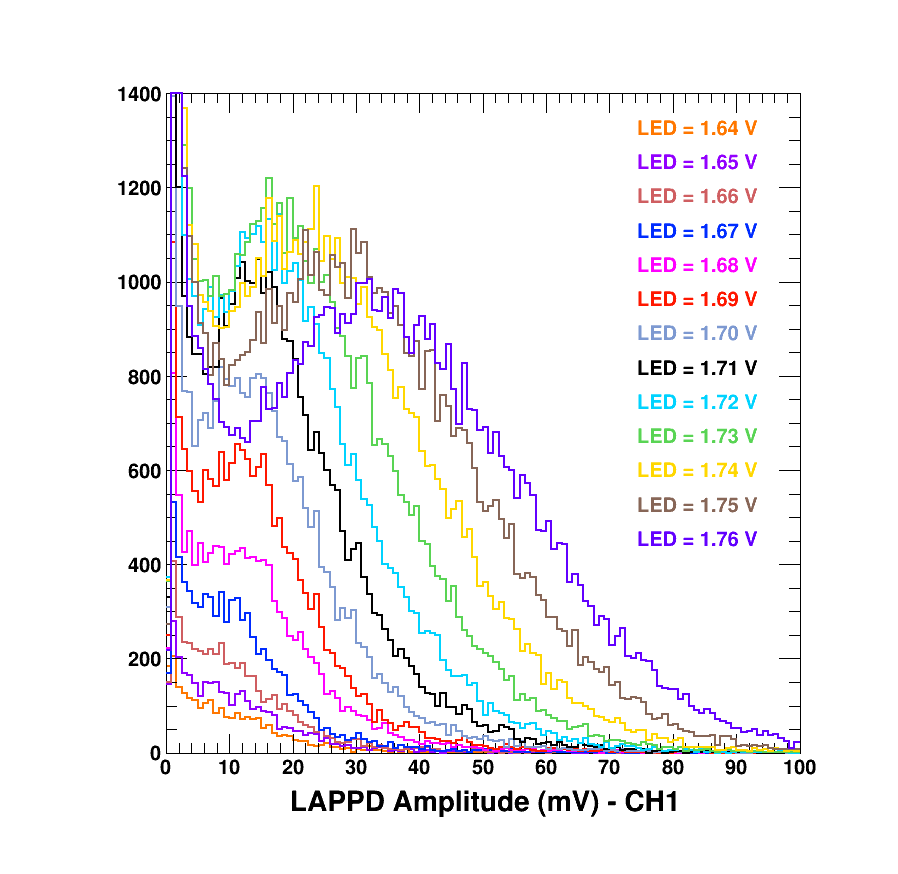}
	\end{subfigure}
        \vspace{-1.5\baselineskip}
	\caption{LAPPD CH1 (CH1 is a label for one uniquely identifiable pixel) signal amplitudes when about 25$\%$ of the LAPPD region corresponding to readout pixel CH1 is illuminated. Top: for clarity only a subset of the data is shown. Bottom: the entire data set is shown.} 
\label{fig:CH1fig}
\end{figure}

\begin{figure}[H]
  \captionsetup[subfigure]{aboveskip=5pt,belowskip=5pt}
	\centering
        \vspace{-1.5\baselineskip}
	\begin{subfigure}{0.78\textwidth} 
		\includegraphics[width=\textwidth]{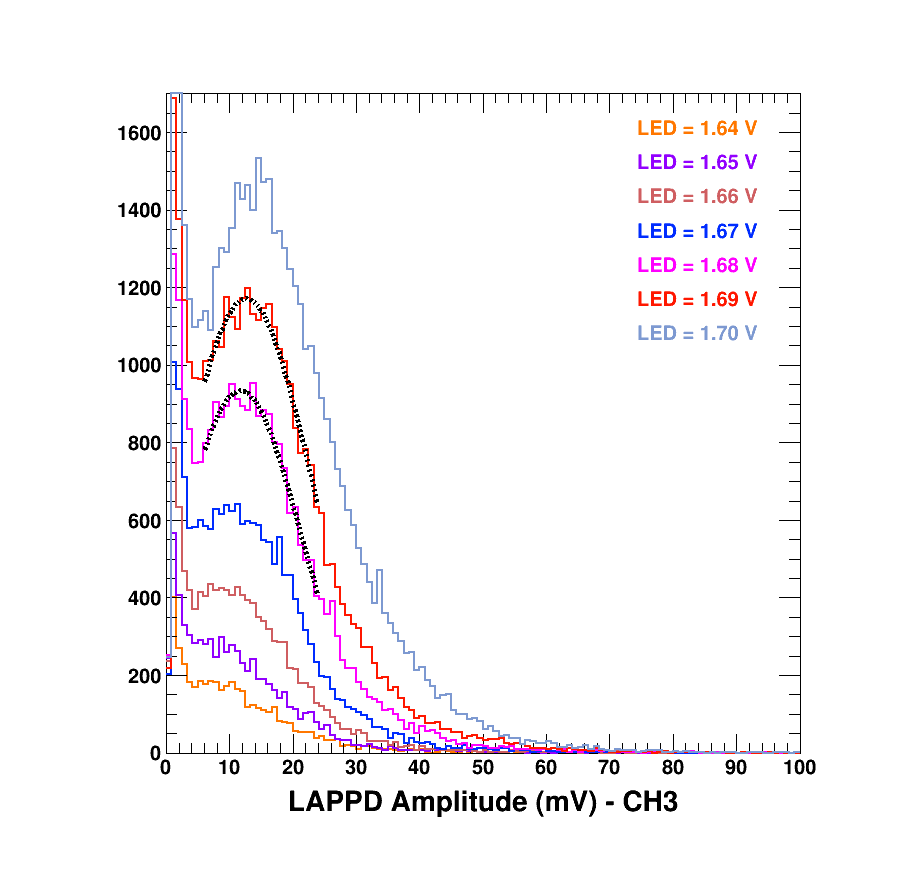}
                \vspace{-3.45\baselineskip}
	\end{subfigure}
	\begin{subfigure}{0.78\textwidth} 
		\includegraphics[width=\textwidth]{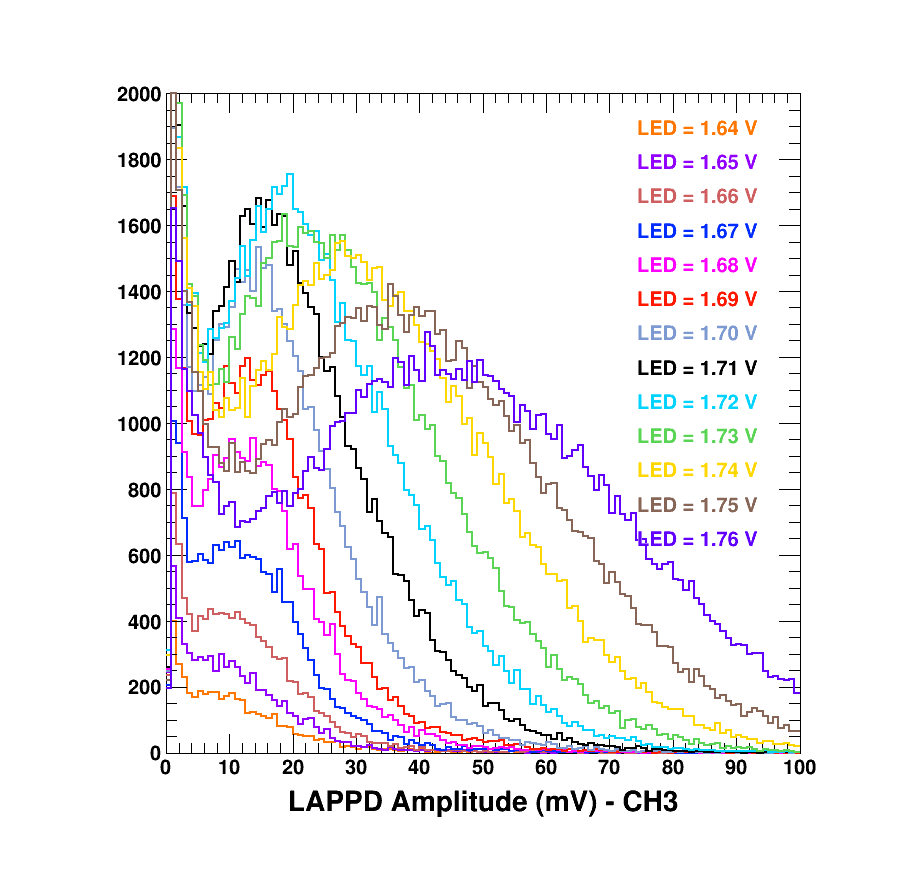}
	\end{subfigure}
        \vspace{-1.5\baselineskip}
	\caption{LAPPD CH3 (CH3 is a label for one uniquely identifiable pixel) signal amplitudes when about 25$\%$ of the LAPPD region corresponding to readout pixel CH3 is illuminated. Top: for clarity only a subset of the data is shown. Bottom: the entire data set is shown.} 
\label{fig:CH3fig}
\end{figure}

\begin{figure}[H]
  \centering
  \includegraphics[width=0.7\textwidth]{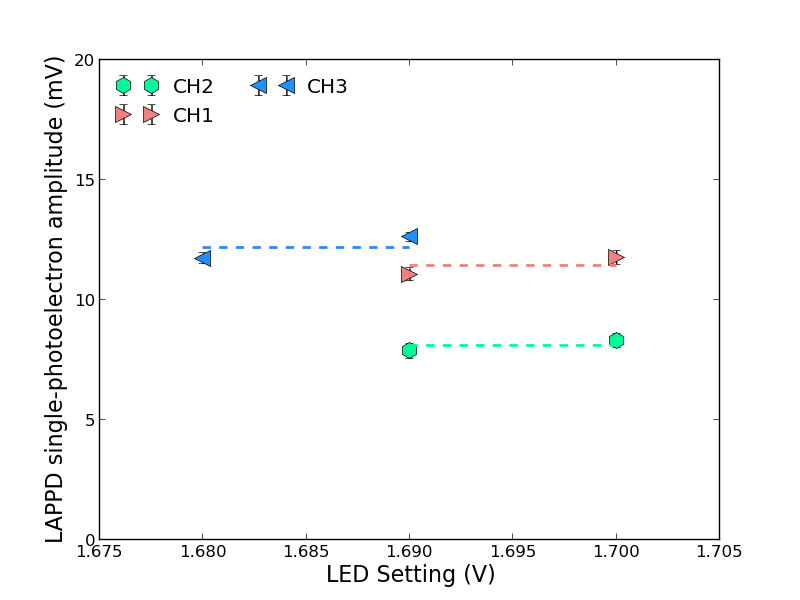}
  \caption{Summary of LAPPD single photoelectron amplitudes from the Gaussian fits.}
  \label{fig:spegain}
\end{figure}

In a second measurement, the LAPPD window was covered so that only 25$\%$ of the monitored pixel was illuminated. This effectively restricted the active area in which photoelectrons could be generated from the photocathode to be smaller than the area of the pixel. The LAPPD pulse amplitude distributions when three individual pixels are thus illuminated one at a time are shown in Figs.~\ref{fig:CH2fig},~\ref{fig:CH1fig} and ~\ref{fig:CH3fig} for LED voltages up to 1.76 V. For clarity, the pulse amplitude distributions obtained with the lowest LED voltages only are shown separately in the top panels of Figs.~\ref{fig:CH2fig},~\ref{fig:CH1fig} and ~\ref{fig:CH3fig}. In the case of this second measurement a very similar pattern to that observed with the maPMT emerges. At the lowest LED voltage settings the LAPPD pulse height distributions for all three pixels monitored tend to be grouped at one average amplitude as single photoelectron production triggers dominate the distributions. As the LED voltage is increased the LAPPD pulse amplitude distributions will contain a mixture of single and multiple photoelectron production triggers. The presence of a valley between the noise signal and the single photoelectron peak is indeed a clear indication of single photoelectron detection. A simple Gaussian fit is performed to roughly determine possible gain changes across the LAPPD and the fit results are shown in Fig.~\ref{fig:spegain}. Gain variations from pixel to pixel up to a factor of two are observed.
  
\begin{figure}[H]
  \centering
 \vspace{-1.5\baselineskip}
  \includegraphics[width=0.75\textwidth]{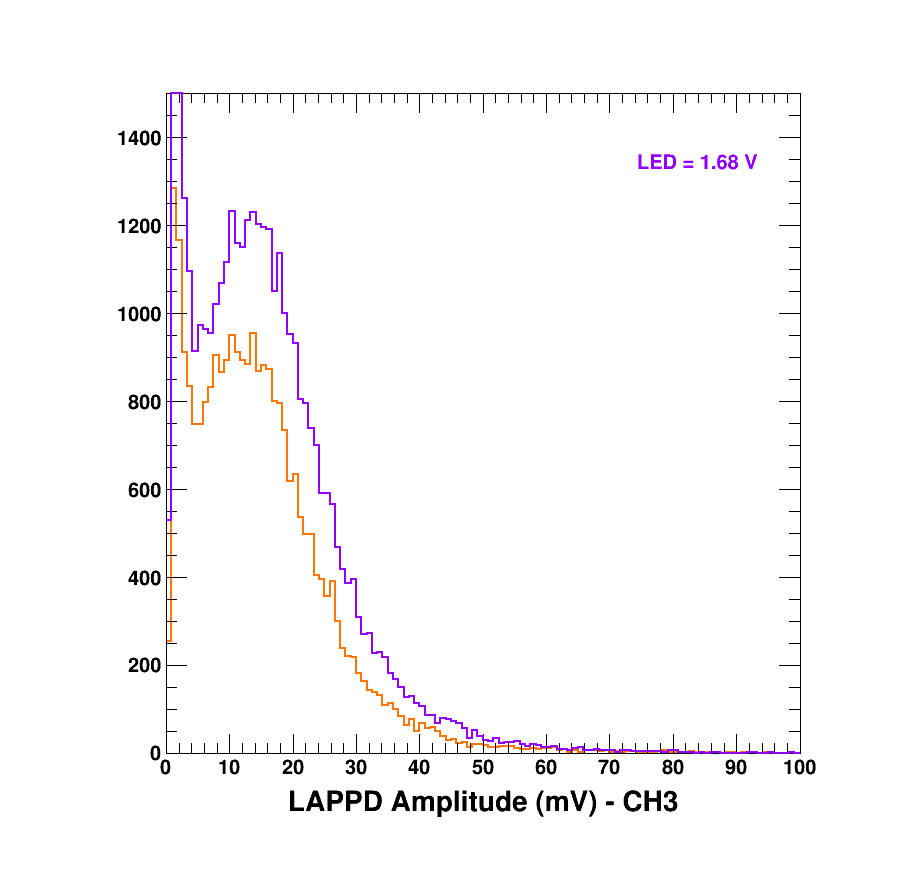}
  \caption{Single photoelectron distributions from LAPPD CH3 are shown for two cases. In orange, only the masked area above CH3 is illuminated. In violet, both the area over CH3 and the area over the neighboring pixel CH1 are illuminated. Masking was used for both pixels.}
  \label{fig:2vs1}
\end{figure}

 In Fig.~\ref{fig:2vs1} single photoelectron amplitude distributions are shown from the same pixel but in one case the adjacent pixel to the one shown was exposed to masked light while in the other case the adjacent pixel was covered. The amplitude distributions have the same average indicating that the masking chosen minimizes the charge spread at the anode to mostly one pixel. Another way used to verify that as a result of the masking the anode charge gets detected on mostly one pixel was by monitoring adjacent pixels to the one illuminated/studied to ensure that a very small amount to no charge was detected.  

\section{Discussion}

The maPMT responded to the small changes in the LED brightness by producing an unchanging pulse amplitude distribution peak over a range of LED voltages. This is indeed consistent with the detection of single photoelectrons. As the LED voltage was further increased, the maPMT average pulse amplitude increased and broadened suggesting the increasing admixture of pulses from one, two or more photoelectrons.
A similar result was obtained with the LAPPD as the LED voltage was increased. However, the pulse height distributions observed at one pixel from illumination of many adjacent pixels were broader than those from the maPMT. This suggests that the single pixel efficiently capacitively coupled to pulses above adjacent pixels, at correspondingly smaller amplitudes. Imposing a mask above one pixel of the LAPPD, so that only 25$\%$ of its area was illuminated produced narrower pulse amplitude distributions as well as the pattern of mostly single photoelectron production events at the lowest LED voltages. 
The effectiveness of the masking was tested by illuminating first one pixel, then both that and the neighboring pixel. The average of the pulse amplitude distributions observed in the two cases were similar, suggesting that very little light and signal was observed from the neighboring pixel.

The LAPPD with its microchannel plates is designed to have nearly continuous spatial sensitivity and consequently position resolution on the order of a millimeter. The maPMT on the other hand, is designed to collect photon signals on discrete anodes, with lower position resolution in trade for lower output electronics burden.
For the purpose of this test the LAPPD photocathode area corresponding to individual readout pixels has been masked primarily to minimize the charge sharing over adjacent readout pixels. A discussion of the performance results reported here must also consider the specific design features of LAPPD 38, and must also consider that these design features can be customized depending on the needs of one application or another.  More specifically, for the capacitively coupled design, the degree to which charge is shared between pixelated anodes, depends upon the spacing between tile components as well as the size and spacing of the anodes. For example, for LAPPD 38 the gaps between photocathode and the top of the first MCP was about 2.8 mm, between MCPs, about 1.1 mm, and between the bottom MCP and the internal ground plane of the detector, about 6.6 mm.  In an application where the goal is minimize charge sharing, and replicate the performance of a multi-anode MCP-PMT, minimizing those gaps would be advantageous.  

Other fundamental design aspects of LAPPDs not explored with our reported test but planned to be investigated in future measurements at Jefferson Lab have to do with LAPPD gain variations. For instance, if two photons strike the photocathode at the same time, depending on the distance between them the photoelectrons extracted may end up getting amplified through pores in the first MCP that are located within a smaller distance than the charge cloud radius per pore at the entrance in the second and subsequent MCPs. In that case the photoelectrons will share pores in MCPs subsequent to the first one and this can lead to the degradation of the pore gain. This effect can be minimized by reducing the distance between MCPs as well as by increasing the voltage between them as means to reduce the cloud charge radius. Thus depending on the photon density and incidence on the photocathode one can select these two parameters (distance and voltage) to produce a LAPPD where pore sharing is minimized or eliminated to prevent active gain variation across the pores. 

Also, any significant intrinsic (passive) gain variation across pores can lead to a poorer resolution for single photoelectron identification since the charge cloud will spread over several pores, at least, in the second and subsequent MCPs. If the Cherenkov detector response is used to form physics triggers, this can be trivially mitigated to a large degree by the use of multichannel amplifiers. Otherwise pixel dependent calibration coefficients can be used in the offline analysis to convert the detector's raw response in yield of photoelectrons per particle detected.

Finally, depending on the readout pixelation and the charge cloud radius at the anode, the charge from one photoelectron after the amplification can spread over more than one readout pixel as discussed above. In that case, if high position resolution is desired one can find the centroid of the charge cloud at the anode by using information from all the pixels that share the charge. If this approach is not desirable and one prefers to minimize the readout pixel sharing, then a readout board with more separation between pixels can be used, and, in addition, one can utilize the distance between the last MCP and anode as well as the voltage between them as knobs to control the charge cloud radius.

\section{Conclusions}

Incom Inc. LAPPD 38 was bench tested at Jefferson Lab to assess its capability of detecting single photoelectron signals. We found that signatures of single photoelectron detection are easily observed when masking of the light impinging on the photocathode is used. It is very encouraging that we clearly detect single photoelectrons with LAPPD 38. The single photoelectron detection resolution when a larger area of the LAPPD is illuminated can be improved by using amplifiers to correct for passive gain variations across the LAPPD. Also, by carefully selecting LAPPD parameters such as distance and voltage between MCPs one can minimize or eliminate pore sharing and prevent active pore gain degradation, if a high density of photons on the photocathode is expected per single trigger. In the near future extensive bench tests of LAPPD 38 are planned at Jefferson Lab to study possible improvements of the single photoelectron detection resolution when a large photocathode area is illuminated as well as to determine possible active gain variations due to pore sharing.  Later on, pending on availability of funds, an in-beam test is planned at Jefferson Lab of a Cherenkov detector prototype with a Incom Inc. LAPPD as a photosensor.

\vspace{0.5cm}
\noindent
\textbf{Acknowledgments}
\vspace{0.4cm}

This material is based upon work supported by the U.S. Department of Energy, Office of Science, Office of Nuclear Physics under contracts DE-AC05-06OR23177. S.P. Malace would like to thank Incom Inc. for providing LAPPD 38 and to the Nuclear Physics Group at Argonne National Lab. She is indebted to Mark A. Popecki and Michael J. Minot from Incom Inc. for very fruitful, insightful discussions of the reported LAPPD test results. She would also like to thank Carl Zorn and Drew Weisenberger from Jefferson Lab for useful discussions.

\end{document}